\documentclass[fleqn,12pt,twoside]{article}
\usepackage{amsmath}
\usepackage{espcrc1}

\usepackage{graphicx}
\usepackage[figuresright]{rotating}

\title{Charge- and parity-projected Hartree-Fock study with the tensor
force of light nuclei\thanks{A part of the present study is
        partially supported by a Grant-in-Aid from
        the Japan Society for the Promotion of Science (14340076).}}

\author{Satoru Sugimoto\address[RIKEN]{Institute of Physical and
Chemical Research (RIKEN),\\ Wako, Saitama 351-0198, Japan},
        Kiyomi Ikeda\addressmark[RIKEN],
    and
    Hiroshi Toki\address{Research Center for Nuclear Physics (RCNP),
    Osaka University,\\ Ibaraki, Osaka 567-0047,
    Japan}\addressmark[RIKEN]
}

\begin{document}
\maketitle
\begin{abstract}
We propose a mean field framework in which the charge and the
parity symmetries of a single-particle state are broken. We break
these symmetries to incorporate the correlation induced by the
tensor force into a nuclear mean field model. We perform the
charge- and the parity projections before variation and obtain a
Hartree-Fock-like equation (charge- and parity-projected
Hartree-Fock equation), which is solved self-consistently. We
apply the Hartree-Fock-like equation to the alpha particle and
find that, by breaking the parity and the charge symmetries, the
correlation induced by the tensor force is obtained.
\end{abstract}
\section{Introduction}
The tensor force, which is mediated by the pion, plays significant
roles in the structure of nuclei. A large part of attractive
energy in light nuclei is due to the tensor force
\cite{akaishi72,akaishi86,pieper01,suzuki98}. Shell model
calculations show that about a half of the single-particle
spin-orbit splittings in $p$- and $sd$-shell nuclei can be
produced by the tensor force \cite{terasawa60}. Furthermore, the
suppression of the effect of the tensor force around and above the
normal density is an essential ingredient of the saturation
mechanism of nuclear matter \cite{bethe71}. These facts should
have a strong impact on the study of the nuclear structure even in
a mean field approach. Therefore, we try to construct a mean field
framework in which the tensor force can be treated on the same
footing as other forces like the central and the LS forces.

To treat the tensor force in a mean field model, we mix positive-
and negative-parity components in a single-particle state (parity
mixing). Further, we mix proton and neutron components in the
single-particle state (charge mixing). As the result, our
single-particle state consists of four components, each of which
has a good parity and a definite charge number. The mixings of
parity and charge is inspired by the specific features of the
pion, which mediates the tensor force. To exploit the pseudoscalar
character of the pion, we introduce the parity mixing, and to take
the isovector character of the pion into account, we introduce the
charge mixing. These are the basic ideas how we treat the pion in
a mean field model. Recently, we applied the parity-mixing mean
field to the relativistic mean field model to include the pion
mean field and showed that the correlation by the pion can be
treated with the parity-mixing mean field \cite{toki02}.

Because a total wave function made of parity- and charge-mixing
single-particle states does not have a good parity and a definite
charge number, the parity and the charge projections need to be
performed to obtain the wave function with a good parity and a
definite charge number. It should be noticed that because the
correlation induced by the tensor force is strong and has a large
effect on a nuclear mean field, these projections should be
performed before variation. Taking the variation of each
single-particle state to minimize the total energy with the
projected total wave function, the charge- and parity-projected
Hartree-Fock (CPPHF) equation for each single-particle state is
obtained. We solve the CPPHF equation for each single-particle
state self-consistently.

In the following, we will explain our formulation briefly and
apply the CPPHF equation to the alpha particle.
\section{Formulation}
We assume a single-particle state with the charge and the parity
mixings in the spherical case as
\begin{align}
 {\psi_{njm}} =
  \sum_{t_z=\pm 1/2}\left(
             {
  \phi_{n l t_z}(r) \mathcal{Y}_{j l m}(\Omega)\zeta(t_z)}
  +{
  \phi_{n \bar{l} t_z}(r) \mathcal{Y}_{j\bar{l} m}(\Omega)\zeta(t_z)}
           \right)
  .
  \label{spwf}
\end{align}
This wave function consists of four terms, each of which has a
good parity and a definite charge number. Here, the radial wave
functions $\phi$'s depend only on the radial coordinate $r$, the
isospin wave functions are denoted as $\zeta(t_z)$ ($t_z=1/2$ for
proton and $t_z=-1/2$ for neutron), and the eigenfunctions of the
total spin $\boldsymbol{j}=\boldsymbol{l}+\boldsymbol{s}$ are
denoted as $\mathcal{Y}$'s. In the spherical case, the good
quantum numbers are $j$ and $m$. $\mathcal{Y}_{j l m}$ and
$\mathcal{Y}_{j \bar{l} m}$ have the same $j$, but different
orbital angular momenta, $|l-\bar{l}|=1$, to implement the parity
mixing.

We assume the form of a total wave function with a good parity,
positive ($+$) or negative ($-$) and a definite charge number $Z$
for $A$-body system as,
\begin{align}
 \Psi^{(\pm;Z)} = \hat{\mathcal{P}}^p(\pm)
 \hat{\mathcal{P}}^{c}(Z) \frac{1}{\sqrt{A!}}\hat{\mathcal{A}} \prod_{a=1}^{A}
 \psi_{n_a j_a m_a},
\end{align}
where $\hat{\mathcal{P}}^p(\pm)$ and $\hat{\mathcal{P}}^{c}(Z)$
are the parity-projection and the charge-projection operators
respectively, and $\hat{\mathcal{A}}$ is the antisymmetrization
operator.

By taking the variation of each single-particle state to minimize
the total energy with the projected wave function,
\begin{align}
\frac{\delta}{\delta \psi_{n_a j_a m_a}} \left\{ \frac{\langle
\Psi^{(\pm;Z)}|\hat{H}
  |\Psi^{(\pm;Z)} \rangle}{\langle  \Psi^{(\pm;Z)} |\Psi^{(\pm;Z)}
 \rangle}
 -
 \sum_{b,c=1}^A \epsilon_{bc}
 \langle \psi_{n_b j_b m_b} |\psi_{n_c j_c m_c} \rangle
 \right\} = 0,
\end{align}
the charge- and parity-projected Hartree-Fock (CPPHF) equation for
$\psi_{n_a j_a m_a}$, which has a Hartree-Fock-equation-like form,
is obtained. In the above equation, $\hat{H}$ is a Hamiltonian of
the $A$-body system. Lagrangian multipliers, $\epsilon_{ab}$ are
introduced to assure the orthonormalization of the single-particle
states, $\langle \psi_{n_a j_a m_a} |\psi_{n_b j_b m_b}
\rangle=\delta_{ab}$. We solve the CPPHF equation
self-consistently with the gradient method.
\section{Application to the alpha particle}
We apply the CPPHF equation to the ground ($0^+$) state of the
alpha particle. As a configuration space, we assume two $j=1/2$
states are filled, each of which accommodates two particles. As
the central force, we take the Volkov No.~1 force \cite{volkov65}
with the multiplying factor $x_\mathrm{TE}$ to the attractive part
of the $^3$E channel. The Majorana parameter is fixed to $0.6$. As
the noncentral force, we take the G3RS force \cite{tamagaki68}
with the multiplying factor $x_\mathrm{T}$ to the isovector
channel of the tensor force, which is a dominant part in the
tensor force and mainly mediated by the pion. We introduce
$x_\mathrm{TE}$ because we treat the tensor force explicitly and
$x_\mathrm{T}$ to check the dependence of the results on the
strength of the tensor force. The factors, $x_\mathrm{TE}$ and
$x_\mathrm{T}$, are determined to reproduce the binding energy of
the alpha particle.

\begin{table}[h]
\caption{Results for the ground ($0^+$) state of the alpha
particle in
 various cases. HF (the second and the third columns) denotes the simple Hartree-Fock scheme.
 PPHF (the fourth column) denotes the
 parity-projected Hartree-Fock scheme in which only the
 parity-projection is performed. CPPHF (the fifth to the last columns) denotes the charge- and
 parity-projected Hartree-Fock scheme in which both the charge and
 the parity projections are performed.
The potential energy ($\langle \hat{v} \rangle$ in MeV), the
kinetic energy ($\langle \hat{T} \rangle$ in MeV), the total
energy ($E$ in MeV), the root-mean-square matter radius
($R_\mathrm{m}$ in fm) and the probability of the $p$-state
component (P(-)) are given. $\langle \hat{v}_\mathrm{C} \rangle$,
$\langle \hat{v}_\mathrm{T} \rangle$, $\langle \hat{v}_\mathrm{LS}
\rangle$, and $\langle \hat{v}_\mathrm{Coul} \rangle$ are the
expectation values for the central, the tensor, the LS, and the
Coulomb potentials, respectively (in MeV). \label{table: cphf}}
\begin{tabular}[t]{crrrrrrrr}
\hline & \multicolumn{1}{c}{HF}&&
 \multicolumn{1}{c}{PPHF}&\multicolumn{5}{c}{CPPHF}\\
$x_\mathrm{T}$&\multicolumn{1}{c}{1.0}&\multicolumn{1}{c}{1.5}&\multicolumn{1}{c}{1.5}&\multicolumn{1}{c}{1.0}&\multicolumn{1}{c}{1.25}&\multicolumn{1}{c}{1.5}&\multicolumn{1}{c}{1.75}&\multicolumn{1}{c}{2.0}\\
$x_\mathrm{TE}$&\multicolumn{1}{c}{1.0}&\multicolumn{1}{c}{0.81}&\multicolumn{1}{c}{0.81}&\multicolumn{1}{c}{0.93}&\multicolumn{1}{c}{0.88}&\multicolumn{1}{c}{0.81}&\multicolumn{1}{c}{0.73}&\multicolumn{1}{c}{0.64}\\
\hline
$\langle \hat{v}_\mathrm{C} \rangle$&      $-$76.67&  $-$56.85&  $-$61.31&  $-$73.60&  $-$70.13&  $-$64.75&  $-$58.34&  $-$50.69\\
$\langle \hat{v}_\mathrm{T} \rangle$&        0.00&    0.00&  $-$10.91&  $-$12.26&  $-$20.23&  $-$30.59&  $-$43.86&  $-$60.41\\
$\langle \hat{v}_\mathrm{LS} \rangle$&     0.00&    0.00&    0.67&    0.75&    1.26&    1.91&    2.72&    3.72\\
$\langle \hat{v}_\mathrm{Coul} \rangle$&   0.83&    0.76&    0.78&    0.85&    0.85&    0.85&    0.86&    0.87\\
$\langle \hat{v} \rangle$&        $-$75.84&  $-$56.10&  $-$70.76&  $-$84.27&  $-$88.26&  $-$92.58&  $-$98.62& $-$106.51\\
$\langle \hat{T} \rangle$&         48.54&   39.98&   49.67&   55.81&   59.71&   64.39&   70.52&   78.19\\
$E$&                              $-$27.30&  $-$16.12&  $-$21.09&  $-$28.46&  $-$28.55&  $-$28.19&  $-$28.10&  $-$28.32\\\
$R_\mathrm{m}$&                                       1.48&    1.63&    1.50&    1.41&    1.39&    1.37&    1.34&    1.31\\
P(-)&                                             0.00&    0.00&    0.08&    0.11&    0.13&    0.16&    0.19&    0.21\\
\hline
\end{tabular}
\end{table}
To see the effect of the parity and the charge mixings, we
calculate three cases. The first calculation is the simple
Hartree-Fock (HF) case, in which we do not perform neither the
parity projection nor the charge projection. The second
calculation is the parity-projected Hartree-Fock (PPHF) case, in
which we only perform the parity projection. The third calculation
is the charge- and parity-projected Hartree-Fock (CPPHF) case, in
which we perform both the parity and the charge projections. We
show the calculated results in Table~\ref{table: cphf}. We fix
$x_\mathrm{T}=1.5$ and $x_\mathrm{TE}=0.81$ for this comparison.
With these parameters the binding energy of the alpha particle is
reproduced in the CPPHF case. We also show the results for the HF
calculation with the original Volkov No.~1 force and the original
G3RS force ($x_\mathrm{T}=1.0$ and $x_\mathrm{TE}=1.0$) as a
reference. In the simple HF case the energy from the tensor force
is zero. It means that the result of the self-consistent
calculation becomes a simple $(0s)^4$ configuration and there is
no $p$-state component. In this case the expectation value of the
tensor force is zero identically, because the tensor force does
not act between $s$ states. If we perform the parity projection
(PPHF), the energy contribution from the tensor force becomes
finite. The kinetic energy becomes larger because some component
of the $s$ states is going up to the $p$ state to gain the
correlation by the tensor force. If we perform the charge
projection further (CPPHF), the contribution from the tensor force
becomes much larger. It is about three times of the one in the
PPHF case. It is reasonable because in the PPHF case only the
$\tau^0_1\tau^0_2$ part in the isovector
($\vec{\tau}_1\cdot\vec{\tau}_2$) channel of the tensor force is
active, but in the CPPHF case all the $\tau^+_1\tau^-_2$, the
$\tau^0_1\tau^0_2$, and the $\tau^-_1\tau^+_2$ parts of the tensor
force are active. These results indicate that both the parity and
the charge projections are very important to treat the tensor
force in the single-particle picture, i.e., a mean field model. We
should note that the variation-after-projection scheme, which we
take here, is necessary because even if we assume the mixings of
parity and charge in the simple Hartree-Fock calculation, we
cannot obtain the result with parity mixing as shown in the third
column in Table 1.

In Table~\ref{table: cphf} we also show the results for various
$x_\mathrm{T}$'s with the CPPHF method to show the dependence of
our results on the strength of the tensor force. If we calculate
the $0^-$ state with the CPPHF scheme using $x_\mathrm{T}$'s and
$x_\mathrm{TE}$'s in Table~\ref{table: cphf}, which reproduce the
binding energy of the ground state of the alpha particle, we
obtain the total energies ranging from $-$2.70 $\sim$ $-$2.83 MeV.

\begin{figure}[hbt]
\centerline{\includegraphics[width=8.4cm]{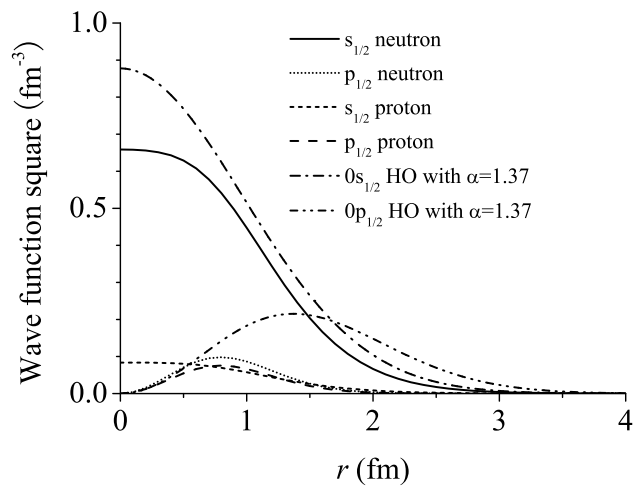}}
 \caption{Wave function squared for the alpha particle with $x_\mathrm{T}=1.5$ and
 $x_\mathrm{TE}=0.81$ as the function of the radial coordinate $r$ (fm).
 The solid curve denotes the $s_{1/2}$ neutron component,
 the dotted curve denotes the $p_{1/2}$ neutron component,
 the short-dashed curve denotes the $s_{1/2}$ proton component, and
 the long-dashed curve denotes the $p_{1/2}$ proton component.
 The harmonic oscillator wave functions with the oscillator length $\alpha=1.37$ (fm) for $0s$ (dashed-and-dotted) and
 $0p$ (dashed-and-double-dotted) are plotted as a reference
 }
 \label{fig: wf}
\end{figure}
In Fig.~\ref{fig: wf}, the squared wave function in the case with
$x_\mathrm{T}=1.5$ and $x_\mathrm{TE}=0.81$ is plotted. We only
show the wave function of the state in which the neutron component
is dominant. In the wave function the probability of the proton
component is 17{\%}. As a reference, the harmonic oscillator wave
functions for $0s$ and $0p$ with the oscillator length
$\alpha=1.37$ (fm), which corresponds to the result with the
original Volkov No. 1 force in the simple Hartree-Fock
calculation. From the figure, you can see that the $p$-state
component mixing into our wave function behaves differently from
the harmonic oscillator $0p$ wave function. It has a much narrower
width. This fact indicates that to obtain the tensor correlation,
a higher-momentum component is necessary to mix.
\section{Summary}
We have developed a mean field framework with parity- and
charge-mixing single-particle states to treat the tensor force. We
showed that by mixing parities and charges in a single-particle
state we can obtain the tensor correlation in a mean field
framework. We found that the projection before variation is
necessary to obtain the tensor correlation. We also found that a
higher-momentum component is needed to gain the energy from the
tensor force.

In this paper, we only treat the alpha particle with spherical
symmetry but it is an interesting subject to calculate heavier
nuclei with our model. The extension to the deformed case is also
interesting. It might give some indication for the relation
between the clusterization in nuclei and the tensor force. We
inevitably treat the short range correlation, which affects the
tensor correlation, to make our model more realistic. Through
these studies, we hope we will reveal the role of the tensor force
in the nuclear structure in the near future.

\end{document}